\documentclass[conference]{IEEEtran}
\IEEEoverridecommandlockouts
\usepackage{cite}
\usepackage{amsmath,amssymb,amsfonts}
\usepackage{algorithmic}
\usepackage{graphicx}
\usepackage{textcomp}
\usepackage{xcolor}
\usepackage{amsmath}
\usepackage{algorithm}
\usepackage{algorithmic}
\usepackage{amsfonts}
\usepackage{graphicx}

\usepackage[english]{babel}

\usepackage{pstool}                    
\usepackage[dvips]{rotating,graphicx}  
\usepackage{float}                     
\usepackage{examplep}

\usepackage{hyperref}

\usepackage{enumitem}  
\usepackage{latexsym}
\usepackage[english]{babel}
\usepackage{csquotes}

\usepackage{siunitx}                        
\usepackage{amsmath}                        
\usepackage{mathtools}                      
\usepackage{bm}  

\usepackage{multirow}       
\usepackage{booktabs}       
\usepackage{makecell}       
\usepackage{tikz}
\usepackage{ragged2e}
\usepackage{xcolor} 
\usepackage{graphicx} 
\usepackage{cite}
\usepackage[nameinlink,capitalize]{cleveref}  

\usepackage{braket}
\usepackage{url} 

\def\BibTeX{{\rm B\kern-.05em{\sc i\kern-.025em b}\kern-.08em
    T\kern-.1667em\lower.7ex\hbox{E}\kern-.125emX}}
\begin{document}

\title{Noise-Aware Distributed Quantum Approximate Optimization Algorithm on Near-term Quantum Hardware
\thanks{Corresponding Author: kuan-cheng.chen17@imperial.ac.uk}
}

\author{
  \IEEEauthorblockN{Kuan-Cheng Chen}
  \IEEEauthorblockA{\textit{QuEST}\\
  \textit{Imperial College London}\\
  London, United Kingdom\\
  kuan-cheng.chen17@imperial.ac.uk}
\and
  \IEEEauthorblockN{  \quad \quad \quad \quad \quad Xiaotian Xu \quad \quad \quad \quad \quad  }
  \IEEEauthorblockA{\textit{QuEST}\\
  \textit{Imperial College London}\\
  London, United Kingdom\\
  xiaotian.xu19@imperial.ac.uk}
\and
\IEEEauthorblockN{ \quad \quad \quad \quad \quad    Felix Burt  \quad \quad \quad  \quad \quad }
  \IEEEauthorblockA{\textit{QuEST}\\
  \textit{Imperial College London}\\
  London, United Kingdom\\
 f.burt23@imperial.ac.uk}
\and
\IEEEauthorblockN{\quad  \quad \quad  \quad  Chen-Yu Liu  \quad \quad \quad  \quad }
  \IEEEauthorblockA{\textit{Graduate Institute of Applied Physics}\\
  \textit{National Taiwan University}\\
  Taipei, Taiwan\\
  d10245003@g.ntu.edu.tw}
\and
\IEEEauthorblockN{\quad  \quad \quad  \quad  \quad \quad  \quad Shang Yu  \quad \quad \quad  \quad  \quad  \quad  \quad }
  \IEEEauthorblockA{\textit{Department of Physics}\\
  \textit{Imperial College London}\\
  London, United Kingdom\\
  shang.yu@imperial.ac.uk}
\and
\IEEEauthorblockN{\quad \quad \quad  \quad   Kin K. Leung \quad  \quad \quad  \quad }
\IEEEauthorblockA{Department of EEE\\
Imperial College London \\
London, UK\\
kin.leung@imperial.ac.uk}}

\maketitle

\begin{abstract}
This paper introduces a noise-aware distributed Quantum Approximate Optimization Algorithm (QAOA) tailored for execution on near-term quantum hardware. Leveraging a distributed framework, we address the limitations of current Noisy Intermediate-Scale Quantum (NISQ) devices, which are hindered by limited qubit counts and high error rates. Our approach decomposes large QAOA problems into smaller subproblems, distributing them across multiple Quantum Processing Units (QPUs) to enhance scalability and performance. The noise-aware strategy incorporates error mitigation techniques to optimize qubit fidelity and gate operations, ensuring reliable quantum computations. We evaluate the efficacy of our framework using the HamilToniQ Benchmarking Toolkit, which quantifies the performance across various quantum hardware configurations. The results demonstrate that our distributed QAOA framework achieves significant improvements in computational speed and accuracy, showcasing its potential to solve complex optimization problems efficiently in the NISQ era. This work sets the stage for advanced algorithmic strategies and practical quantum system enhancements, contributing to the broader goal of achieving quantum advantage.
\end{abstract}

\begin{IEEEkeywords}
Distributed Quantum Computing, Quantum Approximate Optimization Algorithm, Optimization, Compilation
\end{IEEEkeywords}

\section{Introduction}
Quantum computing has emerged as a revolutionary computational paradigm, offering the promise of surpassing classical computing in solving certain computationally intensive tasks. At the heart of QC's potential are the properties of qubits that, through quantum parallelism and the large state space they occupy, allow for significant computational advantages. Pioneering algorithms such as Shor's for integer factorization \cite{shor1994algorithms, shor1999polynomial} and Grover's for database search acceleration \cite{grover1996fast} have laid the foundation for this new era of computation. Recent developments have further demonstrated quantum computing's applicability in various domains, from quantum simulation \cite{cao2019quantum, chen2024quantum} to quantum machine learning \cite{biamonte2017quantum,chen2023quantum}, optimization\cite{zhou2020quantum,liu2022hybrid} , and extending into finance \cite{egger2020quantum} among other fields\cite{chen2020variational,liu2023learning}.

Despite these advances, the dream of scalable, fault-tolerant, universal quantum computing remains a work in progress\cite{shor1996fault}. Today's quantum devices, characterized by the Noisy Intermediate-Scale Quantum (NISQ) era \cite{preskill2018quantum}, are impeded by their modest qubit counts and lack of fault tolerance. It is within this context that the Quantum Approximate Optimization Algorithm (QAOA) \cite{farhi2016quantum} has gained attention as a frontrunner for demonstrating quantum advantage. QAOA's suitability for NISQ devices stems from its requirement for relatively shallow circuits, which align well with the current state of non-error-corrected quantum hardware\cite{pelofske2024short,yanakiev2024dynamic}. Additionally, recent developments in quantum error mitigation and dynamic decoupling techniques, such as IBM Quantum's T-Rex\cite{van2022model} and Zero Noise Extrapolation\cite{giurgica2020digital}, along with Q-CTRL's Automated Deterministic Error Suppression\cite{mundada2023experimental}, have significantly enhanced the fidelity of qubits and the on-device testing performance of quantum algorithms compared to previous achievements\cite{prest2023quantum,chen2023short}.

However, the current landscape of Quantum Processing Units (QPUs) presents significant challenges for the QAOA. The primary obstacles include the limited number of qubits available on contemporary QPUs, constraints on qubit coherence time, qubit fidelity, and the error rates of quantum gates. These factors collectively hinder QAOA's ability to convincingly demonstrate a quantum advantage in solving large-scale complex problems. Pelofske et al. have highlighted these challenges specifically in the context of QAOA applied to Noisy Intermediate-Scale Quantum (NISQ) Computers \cite{pelofske2023high}. Recent approaches such as QAOA-in-QAOA\cite{zhou2023qaoa} and Local-to-Global\cite{yue2023local} attempt to address how to decompose QAOA problems under the constraints of limited qubit numbers, aiming to expand the applicability of QAOA in the NISQ era, for instance, in financial optimization problems. Moreover, Lykov et al. have developed QOKit to facilitate the development of QAOA algorithms through Fast Simulation of High-Depth QAOA Circuits using quantum simulators\cite{lykov2023fast} with NVidia's cuQuantum SDK\cite{bayraktar2023cuquantum}. Additionally, Xu and Chen have developed the HamilToniQ Benchmarking Toolkit\cite{xu2024hamiltoniq}, aimed at benchmarking the performance of QAOA across quantum hardware, protocols, and distributed frameworks. These studies reflect an increasingly comprehensive ecosystem for the QAOA algorithm. This paper introduces a distributed QAOA algorithm based on the concept of noise awareness, aimed at utilizing available quantum resources to achieve more effective scaling and enhance the potential of QAOA to surpass classical optimization methods in the NISQ era.

\section{Background}
In the realm of quantum computing, the Quantum Approximate Optimization Algorithm  (QAOA) is recognized for its potential to address complex optimization challenges, particularly those expressible as cost functions over subsets of the Boolean cube $\mathbb{B}^n$. Central to QAOA is the strategy of approximating solutions to optimization problems—a critical capability in the current era where high-fidelity, large-scale quantum computers remain under development.

The mathematical underpinnings of QAOA involve minimizing a cost function $f: F \rightarrow \mathbb{R}$, represented as a polynomial of spins:

\begin{equation}
f(s) = \sum_{k=1}^{L} w_k \prod_{i \in t_k} s_i, \quad s_i \in \{-1, 1\},
\end{equation}

where $L$ denotes the total number of terms, $w_k$ is the weight of each term, and $t_k$ represents a set of indices, articulating the combinatorial nature of these optimization problems.

To enhance QAOA, our research introduces a distributed computational strategy and a noise-aware approach, optimizing execution on large-scale QPUs and countering quantum noise. This twofold enhancement is pivotal for achieving significant computational speedup and ensuring robust optimization outcomes, even in the presence of noise perturbations.

The effectiveness of QAOA and our enhancements are rigorously tested on two canonical quantum computing problems: the MaxCut problem\cite{guerreschi2019qaoa}, which maximizes the number of edges between partitions of a graph, and the Low Autocorrelation Binary Sequences (LABS) problem\cite{shaydulin2023evidence}, notable for its stringent autocorrelation requirements. By honing the phase ($\gamma$) and mixing ($\beta$) parameter optimization, we guide the quantum system towards an optimal solution state, improving the probability of success upon measurement. The QAOA circuit is enacted through a sequence of phase and mixing operators, governed by the expression:

\begin{equation}
\ket{\vec{\gamma}\vec{\beta}} = \prod_{l=1}^{p} e^{-i\beta_l\hat{M}}e^{-i\gamma_l\hat{C}}\ket{s},
\end{equation}

where $\hat{C}$ is the diagonal Hamiltonian, $\hat{M}$ is the mixer—often chosen as the transverse-field operator $\hat{M} = \sum_i \sigma_i^x$—and $\ket{s}$ is the starting state, typically a uniform superposition. Optimizing $\vec{\gamma}$ and $\vec{\beta}$ aims to minimize the expected solution quality $\bra{\vec{\gamma}\vec{\beta}}\hat{C}\ket{\vec{\gamma}\vec{\beta}}$, leveraging local optimizers.

Our contributions set the stage for addressing real-world large-scale problems in quantum computing, merging advanced algorithmic strategies with practical quantum system enhancements.

\section{Distributed QAOA}

\begin{figure}[htpb]
    \centering
    \includegraphics[width=1\linewidth]{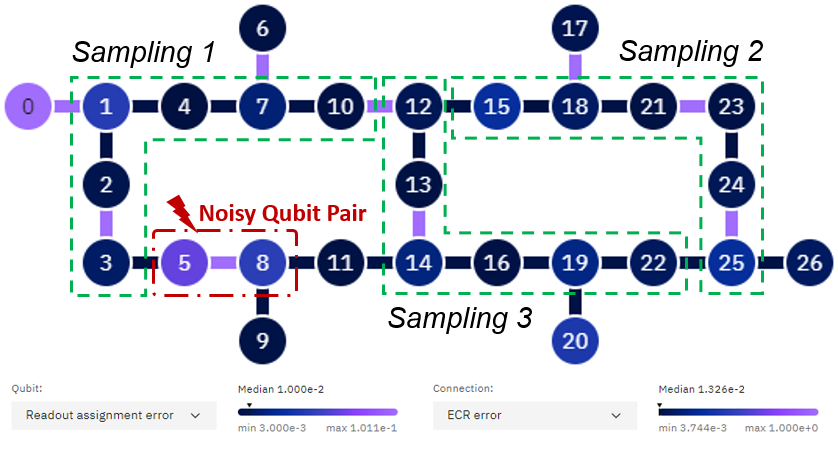}
    \caption{ Visualization of 6-Qubit Distributed QAOA Sampling Process. The diagram depicts three overlapping sampling regions within a 6-qubit network, illustrating the execution of QAOA circuits. Highlighted is a pair of qubits experiencing elevated noise, potentially affecting algorithmic fidelity. The color-coded error metrics denote readout assignment and entanglement crosstalk errors (ECR), with values ranging from the minimum observed error to the maximum for each qubit connection.}
    \label{fig:distri-QAOA}
\end{figure}

\begin{figure*}[htpb]  
    \centering
    \includegraphics[width=\linewidth]{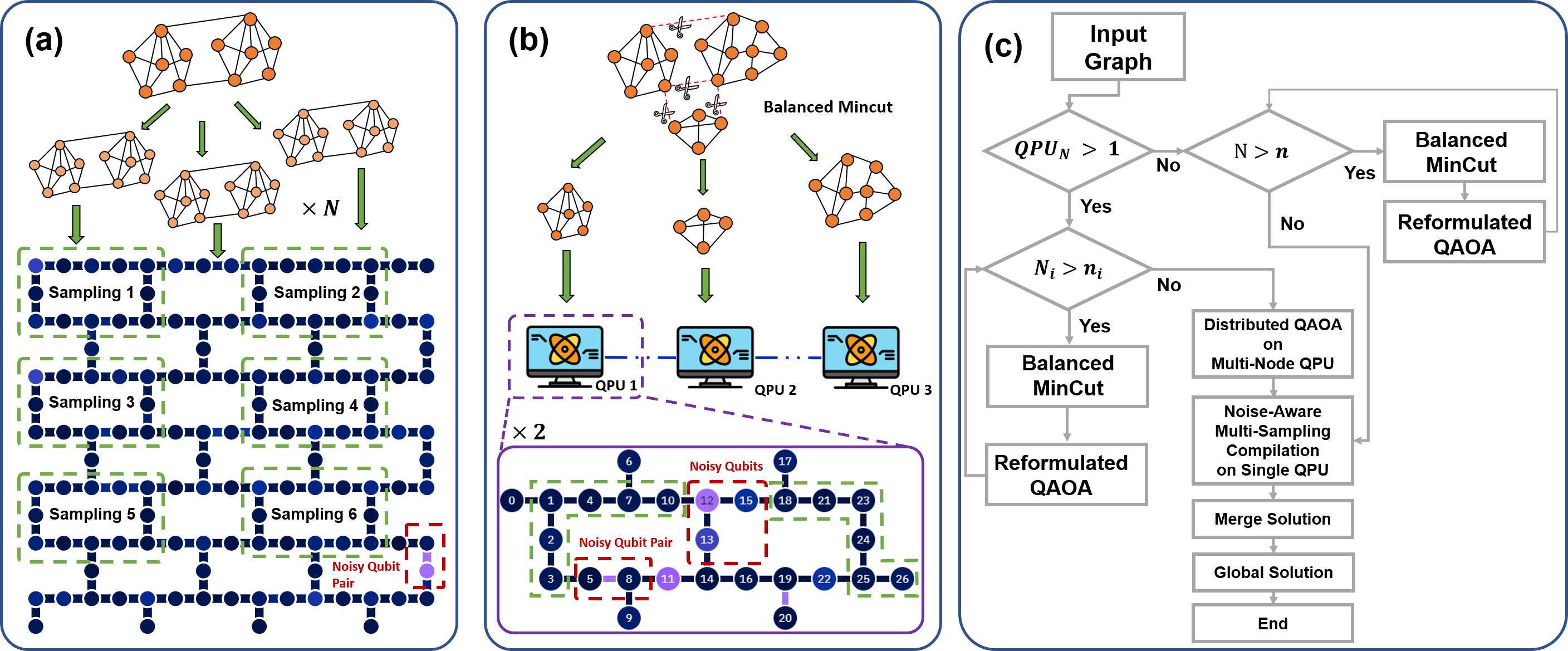}  
    \caption{Integrated Workflow for Noise-Aware Distributed QAOA Execution Across Diverse QPUs. Panel (a) depicts the noise-aware QAOA process on a single-node QPU, utilizing multiple sampling techniques to address and mitigate noise-induced errors. Panel (b) expands the scenario to a distributed quantum computing system, where noise-aware QAOA is applied in conjunction with a balanced MinCut algorithm \cite{g2021efficient} for efficient task distribution across the network. Panel (c) presents a hierarchical approach to solving a MaxCut problem that exceeds qubit capacity; the problem is partitioned by balanced MinCut and progressively reduced and reconstituted into smaller MaxCut instances, aligning with the available qubit resources. The process iterates between multi-node, number of QPU \(QPU_n > 1\), execution as in (b) and single-node operation as in (a), depending on the computational structure and resources at each stage of the problem-solving hierarchy. \(N\) represents the number of qubits required for a complete QAOA problem, while \(N_i\) denotes the number of qubits for the \(i\)-th partitioned QAOA subproblem. Additionally, \(n\) signifies the total number of qubits in a QPU, and \(n_i\) corresponds to the number of qubits in the \(i\)-th QPU within a distributed quantum system. }
    \label{fig:distributed-qaoa-sampling}
\end{figure*}

In Fig. \ref{fig:distri-QAOA}, we present the conceptual diagram for implementing noise-aware distributed QAOA within the framework of distributed quantum computing, as shown in Fig. \ref{fig:distributed-qaoa-sampling}. The core concept of distributed QAOA involves disregarding nodes characterized by low-fidelity qubits and high two-qubit gate errors (greater than $0.01$). Subsequently, the QAOA problem is decomposed into smaller subproblems, which are then compiled onto our QPU using recognized high-fidelity qubits as efficiently as possible. This compilation strategy is both distributed-aware and noise-aware, as discussed in a later section. The compilation demonstrated in this work also employs a series of \texttt{qiskit.transpiler} passes to optimize our circuit for a selected backend, ensuring compatibility with the backend's Instruction Set Architecture (ISA) for each small QAOA subproblem set with a multi-sampling process. This optimization process is facilitated by a preset pass manager from \texttt{qiskit.transpiler}, leveraging its \texttt{optimization\_level} parameter.

The \texttt{optimization\_level} parameter offers varying degrees of optimization\cite{nation2023suppressing}. The lowest level performs only the essential tasks required to execute the circuit on the device, such as mapping circuit qubits to device qubits and inserting swap gates to enable all 2-qubit operations. Conversely, the highest optimization level employs a variety of techniques to significantly reduce the total gate count. Given that multi-qubit gates are prone to high error rates and qubits tend to decohere over time, shorter circuits are likely to yield more accurate results\cite{nation2023suppressing}.

Fig. \ref{fig:distributed-qaoa-sampling}(a) illustrates how distributed QAOA, following the approach described in Fig. \ref{fig:distributed-qaoa-sampling}(c), can achieve linear acceleration by performing multi-sampling on a single-node QPU. Fig. \ref{fig:distributed-qaoa-sampling}(b) further demonstrates how, based on the methodology outlined in Fig. \ref{fig:distributed-qaoa-sampling}(c), distributed QAOA can be scaled up on a distributed quantum computing system for faster, large-scale sampling. This scalability is crucial for realizing the quantum advantage of the QAOA algorithm in the NISQ era.

Algorithm 1 demonstrates the method for decomposing a large-scale QAOA problem into multiple smaller QAOA problems that can be executed on one or more QPUs. If the size of the decomposed QAOA problem is less than twice the number of qubits available on a single QPU, multi-sampling can be employed on a single QPU to accelerate the execution of the QAOA algorithm. This approach achieves a linear speedup in the sampling rate by leveraging multiple samples. The detailed compilation strategy for this process will be discussed in the following section.

\begin{algorithm}
\caption{Noise-Aware Distributed QAOA}
\label{alg
}
\begin{algorithmic}[1]
\REQUIRE Graph $G = (V, E)$, Threshold $\eta$, Node Capacity $n_i$, QPU Capacity $QPU_N$
\ENSURE Global Solution $S$
\STATE \textbf{Procedure} Distributed-QAOA($G$, $\eta$, $n_i$, $QPU_N$):
\IF {$QPU_N > 1$}
\IF {$|V| > n_i$}
\STATE Perform Balanced MinCut on $G$ 
\ELSE
\STATE Perform Reformulated QAOA on $G$
\ENDIF
\ELSE
\IF {$|V| > \eta$}
\STATE Perform Balanced MinCut on $G$ 
\ELSE
\STATE Perform Reformulated QAOA on $G$
\ENDIF
\ENDIF
\IF {$|V| \leq \eta$}
\STATE Execute Distributed QAOA on Multi-Node QPU
\STATE Apply Noise-Aware Multi-Sampling Compilation on Single QPU
\STATE Merge Partial Solutions from QAOA
\STATE Obtain Global Solution $S$
\ENDIF
\RETURN Global Solution $S$
\end{algorithmic}
\end{algorithm}

\section{ Noise-Aware Multi-Sampling Compilation}

Given a QPU with \(N\) qubits, each node in the graph represents a qubit, and each edge represents a two-qubit gate connection. The nodes contain single-readout error rates, and the edges contain two-qubit gate error rates. The compilation strategy aims to optimize the placement of QAOA circuits by selecting qubits and gates with error rates below a specified threshold, \(\eta\). The objective is to ensure reliable quantum computations by ignoring nodes and edges exceeding this error threshold.

\subsection{Step 1: Threshold Filtering}

The initial step in our noise-aware compilation strategy involves defining an error rate threshold $\eta$ for both single-readout error rates and two-qubit gate error rates. Each qubit $q_i$ with a single-readout error rate $e_i$ and each two-qubit gate $g_{ij}$ between qubits $q_i$ and $q_j$ with an error rate $e_{ij}$ are evaluated against this threshold:

\begin{itemize}
    \item $q_i$ is valid if $e_i < \eta$
    \item $g_{ij}$ is valid if $e_{ij} < \eta$
\end{itemize}

Using these criteria, we construct a subgraph $G'$ from the original QPU graph $G$ by retaining only the nodes and edges that meet the error rate criteria:

\[
G' = (Q', E')
\]
\[
Q' = \{ q_i \in Q \mid e_i < \eta \}
\]
\[
E' = \{ g_{ij} \in E \mid e_{ij} < \eta \}
\]

\subsection{Step 2: Noise-Aware Symmetrical Sampling}
For a QAOA problem requiring $n$ qubits, we identify $k$ symmetrical sampling areas. Each sampling area is a subgraph $S_k$ containing $n$ qubits. The number of symmetrical sampling areas $k$ is determined by the number of valid qubits and gates in the filtered subgraph $G'$. To ensure optimal performance, we prioritize sampling areas with the highest fidelity. The fidelity of a block is defined as the product of the readout errors of the selected qubits and the two-qubit gate errors among them:

\[
\text{Fidelity} = \prod_{i=1}^{n} e_i \times \prod_{j=1}^{m} e_{ij}
\]

We evaluate the fidelity for each potential sampling area and select $k$ subgraphs $S_1, S_2, \ldots, S_k$ with the highest fidelity, ensuring each subgraph $S_i$ contains $n$ qubits. Each subgraph $S_i$ maintains symmetry and is topologically equivalent to the other subgraphs.

\subsection{Step 3: Compilation}

For each symmetrical sampling area $S_i$:

\begin{enumerate}
    \item Map the QAOA circuit to the $n$ qubits in $S_i$.
    \item Execute the QAOA circuit on each $S_i$ independently.
    \item Collect results from all sampling areas to improve the robustness and reliability of the computations.
\end{enumerate}

This noise-aware formulation ensures a systematic approach to selecting valid qubits and gates with the highest fidelity, followed by efficient and reliable compilation for QAOA problems on any QPU with $N$ qubits.

\subsubsection{Computational Complexity}
The noise-aware compilation strategy entails several computational steps, each with distinct complexity considerations. The filtering process, which involves checking each qubit and edge against the error rate threshold $\eta$, exhibits a complexity of $O(N + E)$, where $N$ represents the number of qubits and $E$ represents the number of edges. This step is efficient in delineating the feasible set of qubits and connections for further analysis. The identification of subgraphs that meet these criteria depends significantly on the graph's structure and the density of valid subgraphs, with complexity varying based on these characteristics. Finally, the fidelity calculation for potential qubit blocks, crucial for selecting the optimal execution environment for QAOA circuits, has a complexity of $O(N \log N)$. This sorting and selection process is computationally feasible for distributed QPU systems in NISQ era. Each of these steps plays a pivotal role in ensuring that the compilation strategy efficiently narrows down the most reliable computational substrates within quantum hardware constrained by noise and operational imperfections.

\subsection{Evaluation: HamilToniQ Benckmarking Toolkit}
Before introducing the scoring procedure for the compilation strategy of noise-aware distributed QAOA, it is essential to define the criteria by which QPUs are evaluated. In optimization problems, accuracy is the primary focus, making it suitable for the criteria. Instead of using average accuracy for the final H-Score, which diminishes the impact of individual results, HamilToniQ employs the integral of the product between the probability density function (PDF) and a monotonically non-decreasing function\cite{xu2024hamiltoniq}. For instance, consider two individual accuracies $a$ and $b$, with $a > b$. The PDFs of these two results are represented as Dirac delta functions $p_1(x) = \delta(x-a)$ and $p_2(x) = \delta(x-b)$. The shift between these functions along the accuracy axis can be expressed as the variance between two delta functions:
\[
\psi_{a \to b}(x) = \delta(x-b) - \delta(x-a)
\]
By integrating this variance with a monotonically non-decreasing function $h(x)$, we obtain:
\[
\int_0^1 \psi_{a \to b}(x) h(x) \, dx = h(b) - h(a)
\]
This value is non-positive, indicating that $p_1$ shows better performance than $p_2$ when $h(x)$ is increasing, aligning with the fact that $p_1$ corresponds to higher accuracy.

In HamilToniQ, better performance criteria are defined by the collective effect of the above integral. Considering two groups of results, each with $N$ pieces, these groups can be expressed as PDFs $P_1(x)$ and $P_2(x)$. The average variance of the shift effect is given by:
\[
\psi(x) = \frac{1}{N} \sum_{i=1}^N \left[\delta(x-b_i) - \delta(x-a_i)\right] = P_2(x) - P_1(x)
\]
Integrating this variance with a monotonically non-decreasing function $h(x)$ results in:
\[
\int_0^1 \psi(x) h(x) \, dx = \int_0^1 P_2(x) h(x) \, dx - \int_0^1 P_1(x) h(x) \, dx
\]
A non-positive value indicates that $P_1$ outperforms $P_2$. HamilToniQ uses the scoring function derived in the following section to ensure most accuracy values fall within the increasing part of $h(x)$, providing a self-normalized score.

\begin{figure}[htpb]
    \centering
    \includegraphics[width=0.8\linewidth]{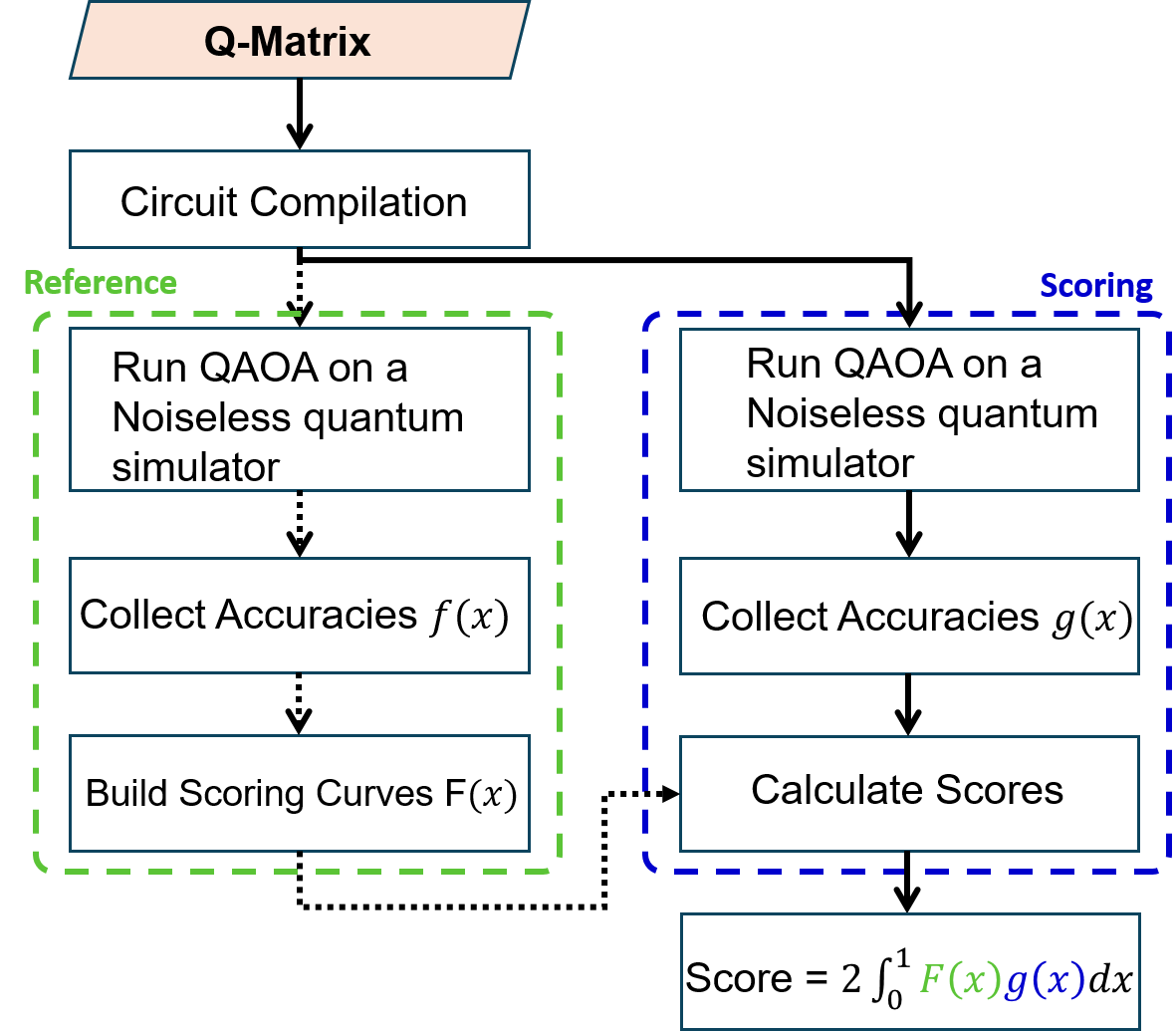}
    \caption{ This flow chart illustrates the standard benchmarking procedures of a QPU (or Distributed QPUs) using HamilToniQ benchmarking toolkit.}
    \label{fig:hamiltoniq}
\end{figure}

In the scoring phase, QAOA is executed $M$ times, and the accuracy $X_i$ from each iteration is used to calculate the score $F(X_i)$. The final H-Score $C$ of a QPU is determined by:
\[
C = \frac{2}{M} \sum_{i=1}^M F(X_i)
\]
When $M$ is large, the distribution of outcomes can be represented as a PDF $g(x)$, and the H-Score becomes the expected value of $F(x)$ over $g(x)$:
\[
C = 2 \int_0^1 F(x) g(x) \, dx
\]
For noiseless QPUs, where the PDF $f(x)$ matches that of a noiseless simulator, the H-Score $C_\text{nl}$ is:
\[
C_\text{nl} = 2 \int_0^1 F(x) f(x) \, dx
\]
Integration by parts shows that $C_\text{nl} = 1$. If a QPU performs better than a noiseless simulator, the H-Score can exceed $1$, reaching up to $2$ in ideal cases. This benchmarking scheme, illustrated in Fig. \ref{fig:hamiltoniq}, highlights the effectiveness of QPUs in real-world optimization scenarios.

\section{Result}
\subsection{Multi-Sampling on Multi-QPUs}
\begin{figure}[htpb]
    \centering
    \includegraphics[width=\linewidth]{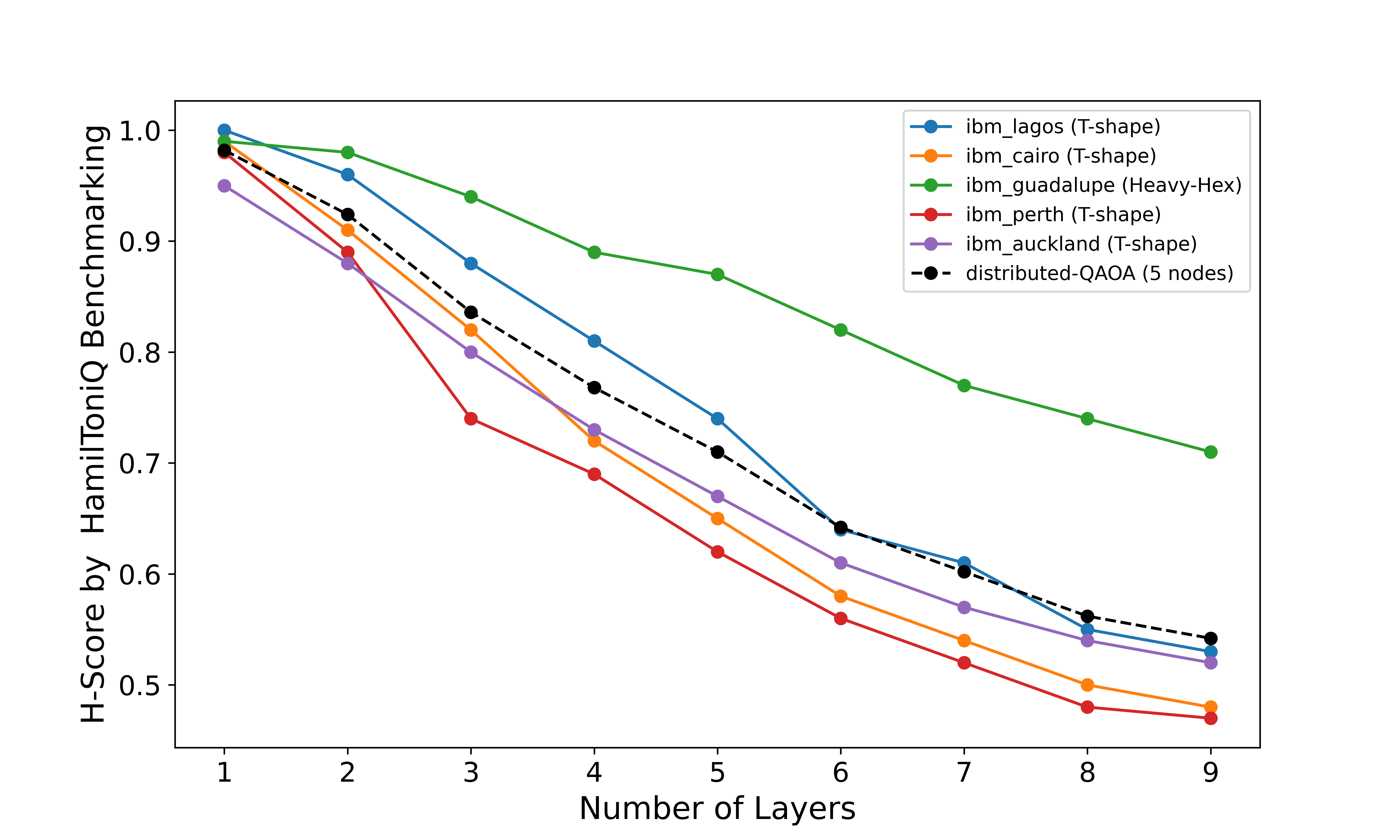}
    \caption{Benchmarking H-Score by the HamilToniQ Toolkit across various quantum devices with different topologies. The devices include IBM Lagos, IBM Cairo, IBM Perth, IBM Auckland (T-shape), and IBM Guadalupe (Heavy-Hex), contrasted with distributed-QAOA (5 nodes). The H-Score, indicative of quantum fidelity, is plotted against the number of layers within the quantum circuit. The graph illustrates the relative performance of each device topology, with distributed-QAOA showcasing enhanced sampling efficiency in the distributed quantum system.
}
    \label{fig:score}
\end{figure}

HamilToniQ utilizes High-Performance Computing and the cuQuantum SDK\cite{bayraktar2023cuquantum} to simulate a noiseless quantum simulator for solving the distribution results of QAOA. These results then serve as a reference for evaluating the performance of QPUs, as well as the effectiveness of protocols and architectures. Within our distributed QAOA framework, we can achieve linear acceleration even on distributed quantum systems with small QPUs, as shown in Fig. \ref{fig:score}. We utilize five QPUs as nodes to accomplish six samplings for a 6-qubit QAOA problem (where \textit{ibm\_guadalupe}, with 16 qubits, can accommodate two sampling zones on a single chip). Thus, we can faster solve problems through sampling on various available QPUs under controllable H-Scores.

\subsection{Distributed QAOA on Multi-QPUs}

\begin{figure}[!b]
    \centering
    \includegraphics[width=0.9\linewidth]{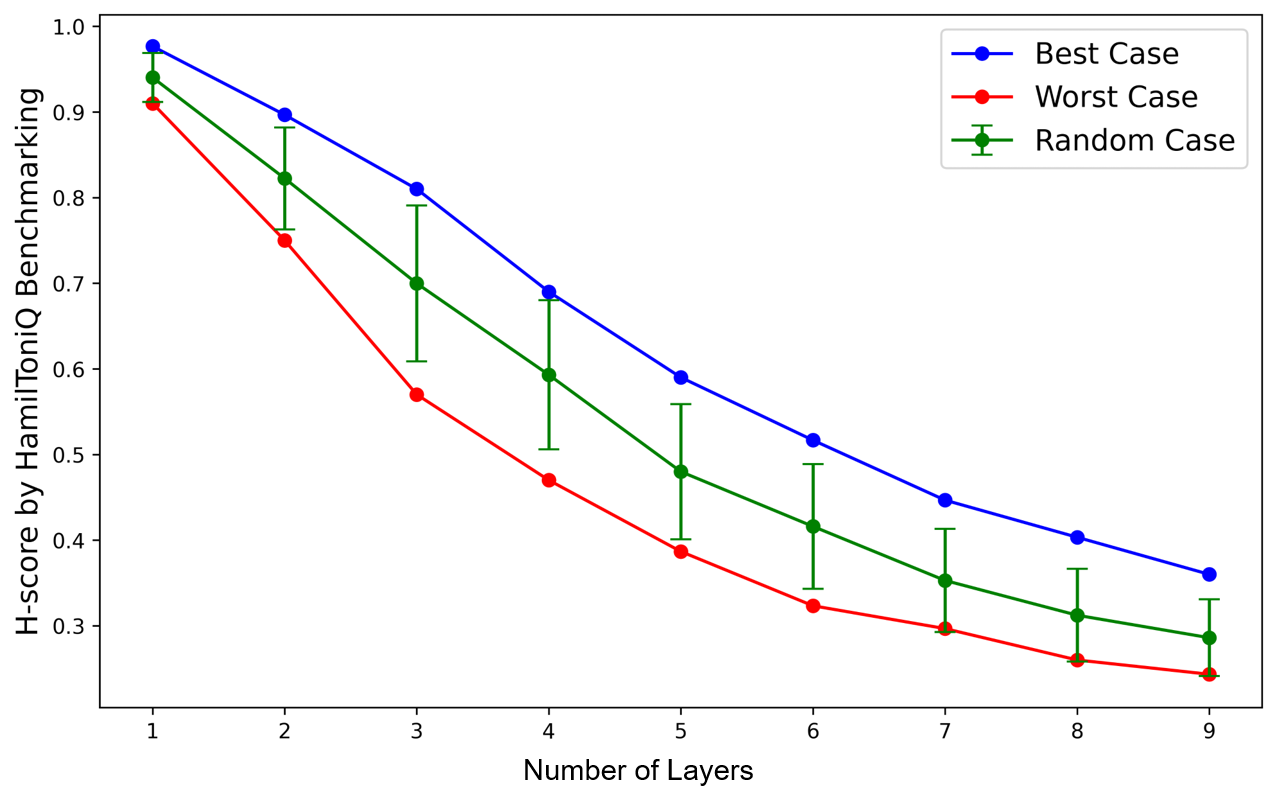}
    \caption{Validation and comparison of H-scores for best, worst, and random QPU selection across layers in a distributed QAOA benchmarking.
}
    \label{fig:dist-score}
\end{figure}

In this study, we also explore the potential of noise-aware distributed QAOA processing on multiple QPUs. We demonstrate this by addressing a 15-qubit QAOA problem, which can be decomposed into smaller QAOA problems of 4, 5, and 6 qubits using the Balanced MinCut method \cite{g2021efficient}. We utilize the results from HamilToniQ benchmarking as a resource management toolkit, prioritizing QPUs with higher overall scores to handle the QAOA subproblems with larger qubit counts\cite{xu2024hamiltoniq}. Fig. \ref{fig:dist-score} illustrates the differences in H-scores between the best and worst IBM Quantum QPUs from our benchmarking study. The use of HamilToniQ as a resource management toolkit significantly enhances the accuracy of large-scale QAOA algorithms by efficiently allocating computational resources.

\section{Conclusion}
In this paper, we propose a noise-aware distributed QAOA framework for real-hardware compilation strategy. This framework is capable of conducting rational QAOA computational distribution based on known hardware calibration data (readout error, two-qubit-gate error). It enables QAOA to utilize limited quantum resources more efficiently to achieve faster sampling (a greater number of shots) and to more effectively allocate QAOA-within-QAOA problems, thus efficiently solving large-scale QAOA issues. Our proof-of-concept simulations have demonstrated the potential of the proposed framework in constructing a distributed quantum computing framework for large QAOA algorithms and have compared it with the methods of a naive QAOA strategy. Moreover, our versatile framework can be applied in conjunction with other distributed compilation strategies or quantum error mitigation techniques to explore more effective compilations of QAOA under the limited quantum resources of the NISQ era, aiming to solve more complex optimization problems.

\section{Acknowledgements}
This work was supported by the U.K. Engineering and Physical Sciences Research Council (Grant No. EP/W032643/1).


\bibliographystyle{siamurl}
\bibliography{references}

\end{document}